\documentclass[apl, twocolumn, superscriptaddress]{revtex4}

\usepackage{graphics}
\usepackage{subfigure}
\usepackage{graphicx}
\usepackage{color}
\usepackage{xspace}
\usepackage{longtable}

\def\deg{^\circ\xspace}
\def\degC{$^\circ$C\xspace}
\def\etal{\emph{et al.}\xspace}

\newcommand{\twobyone}{\ensuremath{\mathrm{2\times1}}\xspace}

\newcommand{\ctwobytwo}{\ensuremath{\mathrm{c(2\times2)}}\xspace}

\newcommand{\sioneoneo}{$[\bar{1}\bar{1}0]$/$[1\bar{1}0]$\xspace}
\newcommand{\sioneoneone}{$[\bar{1}\bar{1}\bar{1}]$/$[2\bar{1}\bar{1}]$\xspace}

\def\degC{$^\circ$C\xspace}
\def\etal{{\em et al.\ }}

\begin{document}

\title{Medium-Energy Ion Scattering study of strained holmium silicide nanoislands grown on silicon (100)}
\date{\today}

\author{T.J. Wood}
\affiliation{Department of Physics, University of York,
Heslington, York, YO10 5DD, UK}
\author{C. Eames}
\affiliation{Department of Physics, University of York,
Heslington, York, YO10 5DD, UK}
\author{C. Bonet}
\affiliation{Department of Physics, University of York,
Heslington, York, YO10 5DD, UK}
\author{M. B. Reakes}
\affiliation{Department of Physics, University of York,
Heslington, York, YO10 5DD, UK}
\author{T.C.Q. Noakes}
\affiliation{STFC Daresbury Laboratory, Daresbury, Warrington, WA4
4AD, UK}
\author{P. Bailey}
\affiliation{STFC Daresbury Laboratory, Daresbury, Warrington, WA4
4AD, UK}
\author{S.P. Tear}
\email[e-mail: ]{spt1@york.ac.uk} \affiliation{Department of
Physics, University of York, Heslington, York, YO10 5DD, UK}

\begin{abstract}
We have used medium-energy ion scattering (MEIS) to quantitatively
analyse the structure of holmium silicide islands grown on the
Si(100) surface. Structure fitting to the experimental data
unambiguously shows that the tetragonal silicide phase is present
and not the hexagonal phase which is associated with the growth of
nanowires at submonolayer coverages. Islands formed with a lower
holmium coverage of 3 ML are also shown to be tetragonal which
suggests that the hexagonal structure is not a low coverage
precursor to the growth of the tetragonal phase. MEIS simulations
of large nanoislands which include the effects of lateral strain
relief have been performed and these compare well with the
experimental data.
\end{abstract}

\maketitle

\begin{figure}
\centering
    \subfigure[]{
    \label{fig: latticematch}
    \includegraphics[width=3.0in]{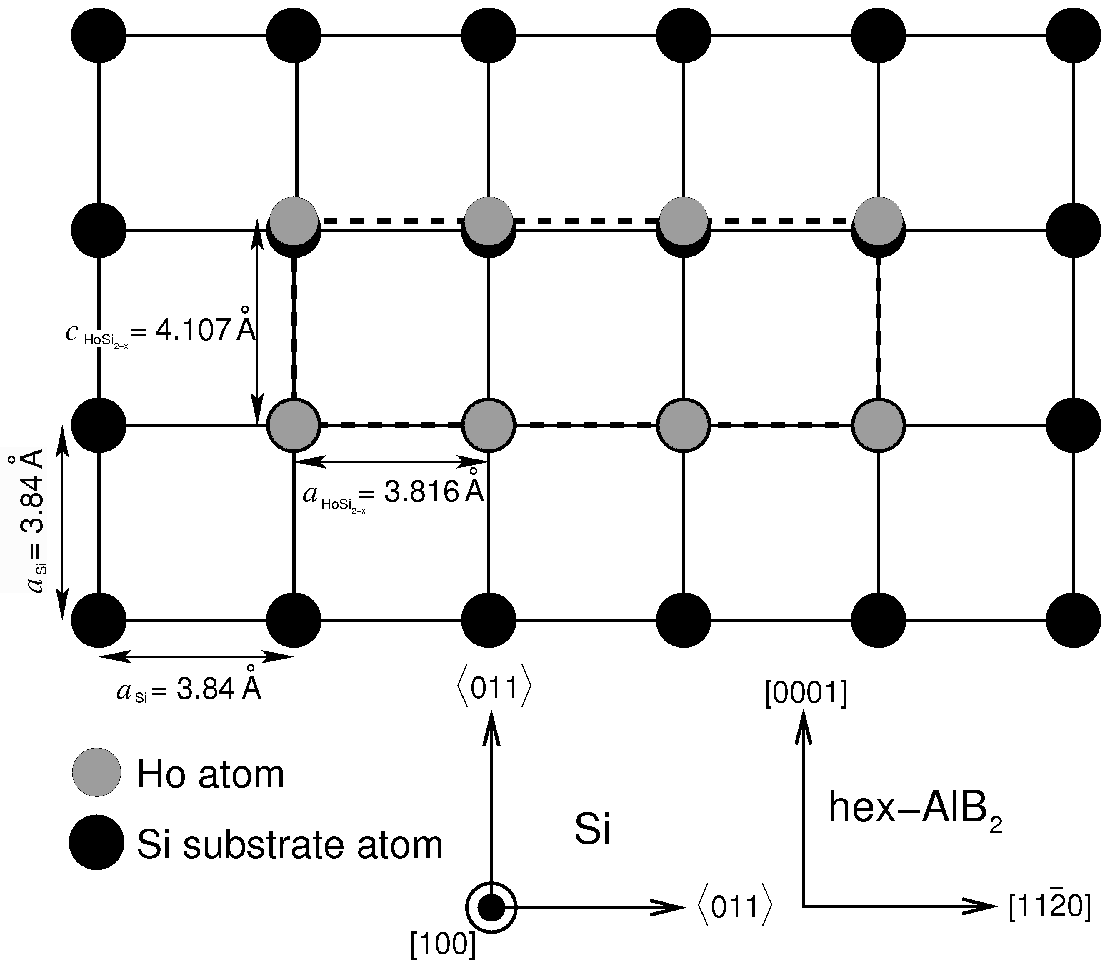}}
    \subfigure[]{
    \label{fig: hexagonal}
    \includegraphics[width=3.7in]{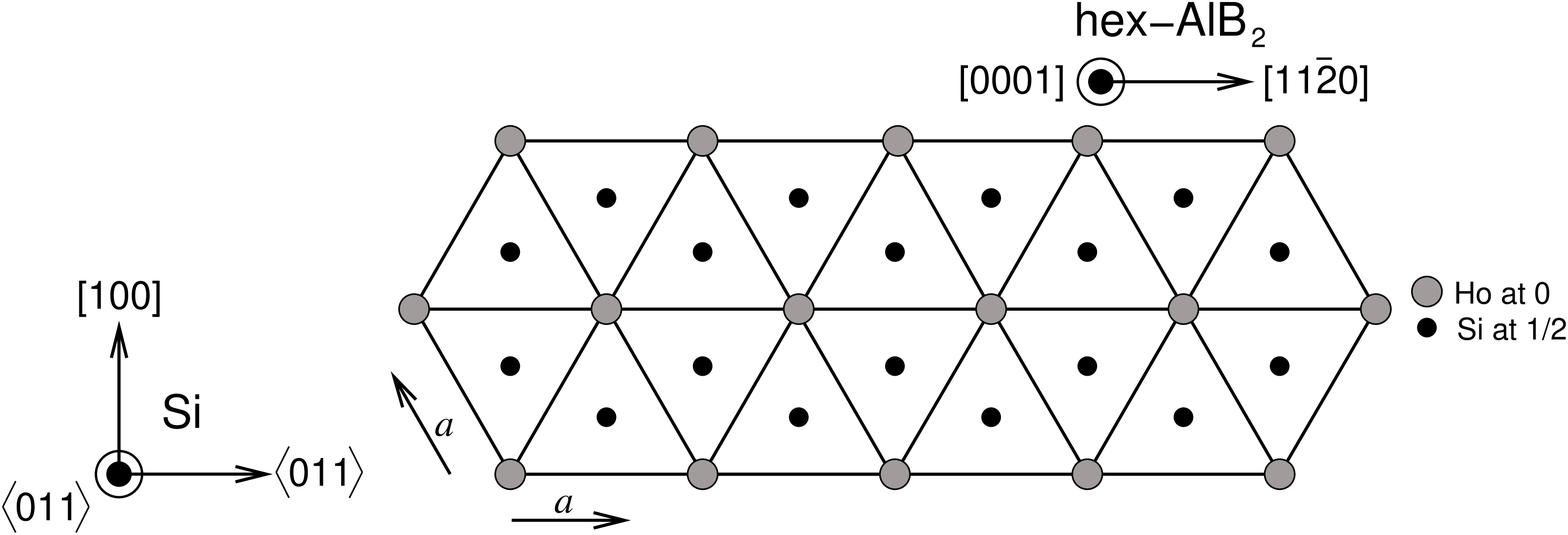}}
    \caption{(a) A top down schematic illustrating the lattice
    matching of the hexagonal defect-AlB$_2$ structure to the
    Si(100) surface. The anisotropy in lattice mismatch creates the
    high aspect ratio nanowires commonly observed for several RE
    metals. (b) A side view schematic of the AlB$_2$ structure.
    }
    \label{fig: hexagonalstructure}
\end{figure}

\noindent Whilst there have been many studies of rare-earth (RE)
silicides on the Si(111) surface, relatively little was known
about the growth mechanisms on the Si(100) surface until the
discovery of self-assembled nanowires by Preinesberger \etal
\cite{Preinesberger1998A} These novel structures form when a
suitable RE metal, eg.\ Gd, \cite{Kirakosian2002A, McChesney2002A,
Chen2002C, Lee2003A, Liu2003B} Sc, \cite{Chen2002C} Dy,
\cite{Preinesberger1998A, Chen2003A, Nogami2001A, Liu2001A,
Liu2003A} Sm, \cite{Chen2002C} Er, \cite{Chen2003A, Nogami2001A,
Chen2000A, Chen2002A, Cai2004A} Ho \cite{Nogami2001A,
Ohbuchi2002A} or Y, \cite{Katkov2002A} is deposited onto a clean
Si(100) substrate held at an elevated temperature. Characteristic
`wires' measuring up to a micrometre in length, and typically only
a few nanometres wide were observed by scanning tunneling
microscopy (STM). This discovery and the potential technological
applications of such conducting nanowires has motivated
considerable interest over the last ten years \cite{Frangis1996A,
Frangis1997A, Chen1998A, Chen2000A, Nogami2001A, Liu2001A,
Chen2002A, Ohbuchi2002A, Chen2002B, Kirakosian2002A,
McChesney2002A, Katkov2002A, Chen2002C, Chi2003A, Liu2003B,
Liu2003A, Lee2003A, Chen2003A, Kuzmin2004A, Cai2004A, He2004A,
Harrison2005A, Peto2005A, Tsai2005A, Pasquali2005A, Ye2006A,
Ye2006B}. Chen \etal have demonstrated that the growth of
nanowires with extremely high aspect ratios is the result of the
anisotropic lattice mismatch that results from the growth of the
hexagonal, defect-AlB$_2$ RE silicide on the Si(100) surface (see
Fig.~\ref{fig: hexagonalstructure}) \cite{Chen2000A}.

\begin{figure}
  \begin{center}
  \includegraphics[width=1.5in]{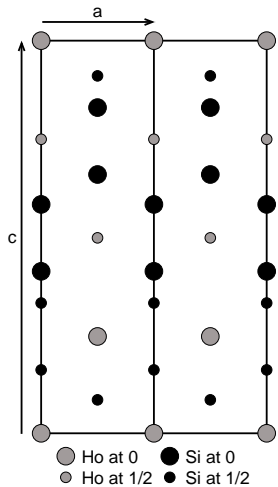}
  \caption{A side view schematic of the tetragonal ThSi$_2$
  structure. The orthorhombic GdSi$_2$ structure of HoSi$_2$ is essentially the
  same but with $a =$ 4.03 \AA \,and $b =$ 3.94 \AA, a difference of just 2.25 \%.}
  \label{fig: tetragonal}
  \end{center}
\end{figure}

Depositing more RE metal causes islands to form with the surface
displaying a \ctwobytwo periodicity. The structure of these RE
silicide islands has been proposed to be either hexagonal,
tetragonal (ThSi$_2$) or orthorhombic (GdSi$_2$)
\cite{Preinesberger1998A, Frangis1996A, Frangis1997A, Chen2002A,
Liu2003A, Liu2003B, Cai2004A, He2004A, Ye2006A, Ye2006B}. All
three of these phases are known to exist in the bulk and their
lattice constants have been measured and it is known that the bulk
orthorhombic phase is only a small distortion of the tetragonal
phase \cite{metalsilicides}. A side-view schematic of the ThSi$_2$
structure is shown in Fig.~\ref{fig: tetragonal}. Evidence for the
growth of the tetragonal form has generally been inferred from
lattice mismatch arguments, along with STM measurements of step
heights in the silicides on the Si(100) surface \cite{Cai2004A}.
SXRD measurements also yield information which appears to confirm
that Er silicide islands adopt the tetragonal form, with the
surface \ctwobytwo periodicity interpreted as being due to Si
adatoms \cite{Chen2002A}.

STM experiments reveal that the island morphology is very
sensitive to the RE metal deposited and to the annealing
temperature used. Those RE metals (Nd, Sm, Yb) that have a low
anisotropy in their lattice match to the substrate form compact 3D
islands \cite{Nogami2001A}. Those that have a high anisotropy form
both elongated and compact islands, depending upon the growth
temperature. Growth at lower temperatures (600$^{\circ}$C) causes
the formation of elongated islands and growth at temperatures
above 650$^{\circ}$C results in compact 3D islands. However, the
two island morphoplogies have been observed to coexist, especially
when using intermediate annealing conditions.

Dysprosium silicide provides an interesting case study. The
elongated islands have typical dimensions 2 nm high, 15 nm wide,
and 500 nm long and the smaller compact islands have typical
dimensions 5 nm high, 50 nm wide and 200nm long \cite{He2004A}.
Using high resolution cross-sectional TEM, Ye \etal demonstrated
that elongated and non-elongated islands coexist when Dy silicide
is grown at 600 \degC $-$ 650 \degC on the Si(100) surface
\cite{Ye2006A}. They proposed that the structure of the elongated
islands was hexagonal, and the excessive stress within this
structure (due to the large $c$-axis mismatch) is relieved through
dislocations and tilting across the width of the island. The
non-elongated islands were found to be a tetragonal or
orthorhombic structure with only a small amount of tilting
required to relieve the stress, since the lattice match is
relatively small in both directions. It was also noted that there
was some expansion of the $c$-axis to relieve the stress in much
the same way as has been demonstrated for the 2D and 3D RE
silicides on Si(111) \cite{Bonet2005A,Wood2006A}. However, the
recent HR-TEM study by He \etal claims that the compact 3D Dy
silicide islands are in fact a fully relaxed, and hence stress
free, hexagonal form, whilst the elongated nanowire islands are
tetragonal/orthorhombic, with the faulted stacking relieving the
stress in the structure \cite{He2004A}. The delicate energy
balance within the silicide in this particular example is also
highlighted by the fact that there is both a tetragonal and an
orthorhombic phase of Dy silicide in the bulk.

Despite the many studies that have been conducted, it has been
noted in recent work that no ideal crystallographic determination
has been possible to clearly prove once and for all the true
nature of the 3D islands \cite{Pasquali2005A, He2004A, Ye2006B}
created when the rare earth metals form a silicide in this way.
MEIS is of particular value in the study of these nanostructures
since the large mass contrast of Ho and Si allows the elucidation
of the silicide structure free from substrate effects. This also
means that the technique is able to isolate the regions of
interest on the surface by only selecting the Ho signal for the
structural optimisation.

\section{EXPERIMENT}

All MEIS data were obtained at the UK MEIS facility at STFC
Daresbury Laboratory. The experiments were conducted under
ultra-high vacuum (UHV), with typical base pressures of
$2~\times~10^{-10}$~mbar. Clean Si(100) samples (cut from a
lightly doped n-type wafer) were prepared by {\em e}-beam heating
to $\sim$~$1200$~\degC for 1~min, then slowly cooled
($<$~100~\degC~min$^{-1}$) between 1000~\degC and 600~\degC to
ensure an ordered surface was obtained. Temperature measurements
were made using an infrared pyrometer. The heating cycles were
repeated until a sharp \twobyone low-energy electron diffraction
(LEED) pattern was observed. The 3D silicide islands were then
formed by depositing approximately 6~ML of Ho from a quartz
crystal calibrated tantalum boat source onto this clean Si(100)
surface, which was held at $\sim650$~\degC during deposition and
for 5~min afterwards. When a \ctwobytwo LEED pattern was observed
and the Auger electron spectrum showed the samples to be free from
contamination, they were transferred under UHV into the ion
scattering chamber.

\begin{figure}
\begin{center}
\includegraphics[width=2in]{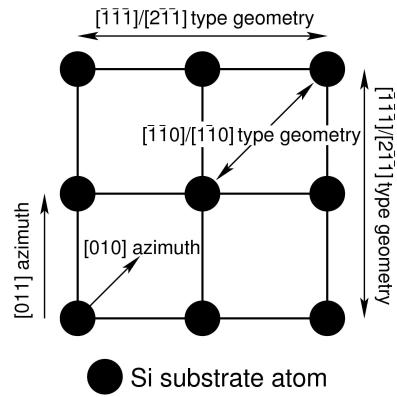}
\caption{View from above of the bulk terminated Si(100) surface
showing the two incident beam directions used in the MEIS
experiments. For the \sioneoneo geometry the angle with respect to
the sample normal was 45$^{\circ}$ and for the \sioneoneone
geometry the polar angle was 54.74$^{\circ}$. The scattering plane
through the tetragonal silicide is shown for the \sioneoneo
geometry in figure~\ref{fig: scattering_plane}.} \label{fig:
geometries}
\end{center}
\end{figure}

The ion scattering data were taken using 100 keV H$^{+}$ ions
incident upon the sample and the scattered ions were detected
using an angle-resolving toroidal-sector electrostatic ion-energy
analyser and its microchannel plate detector. The MEIS spectra
confirmed that the samples were free of contaminants and data were
acquired with a total dose of $\sim 10^{16}$ ions cm$^{-2}$.
Further details about the Daresbury MEIS facility can be found in
the literature
\cite{Spence2000A,Spence2002A,Spence2000B,Bonet2005A,Wood2005A}.
During each experiment two different incident beam directions onto
the sample were used. These beam directions are shown with respect
to the substrate in fig.~\ref{fig: geometries}. The notation
$[in]/[out]$ for each geometry defines the ingoing crystal
direction $[in]$ and an outgoing crystal direction $[out]$ that
lies in the detected angular range of the scattered ions.

\section{STRUCTURE FITTING: TETRAGONAL OR HEXAGONAL?}

Due to the contention in the literature regarding whether these
islands take the hexagonal, tetragonal or orthorhombic form, MEIS
proves particularly powerful since it can clearly demonstrate the
presence of one of these structures. In this work we have used the
XVegas code \cite{Frenken1986A} which uses Monte-Carlo methods to
simulate the blocking curves of a proposed structural model.

\begin{figure}[h]
\centering
    \subfigure[]{
    \label{fig: hexlayers}
    \includegraphics[width=2.9in]{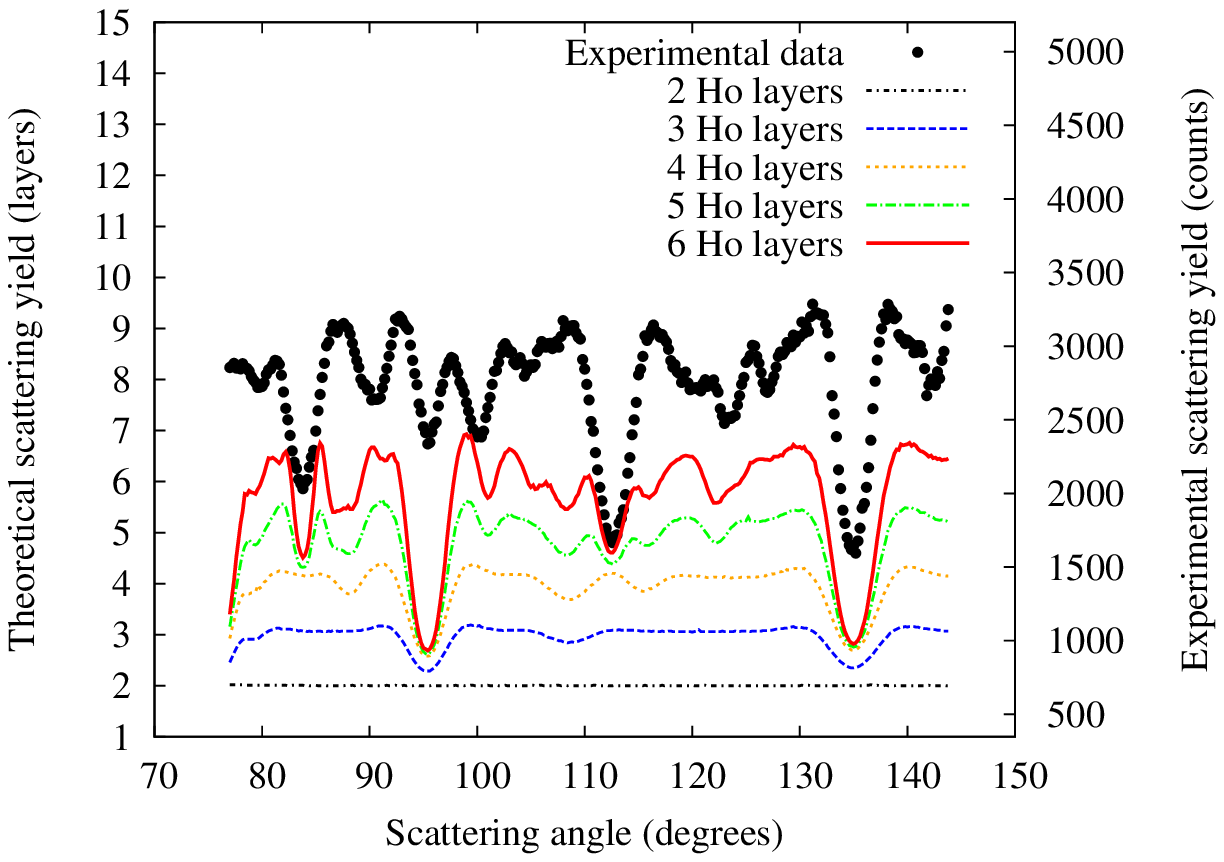}}
    \subfigure[]{
    \label{fig: hex6layerfit}
    \includegraphics[width=2.9in]{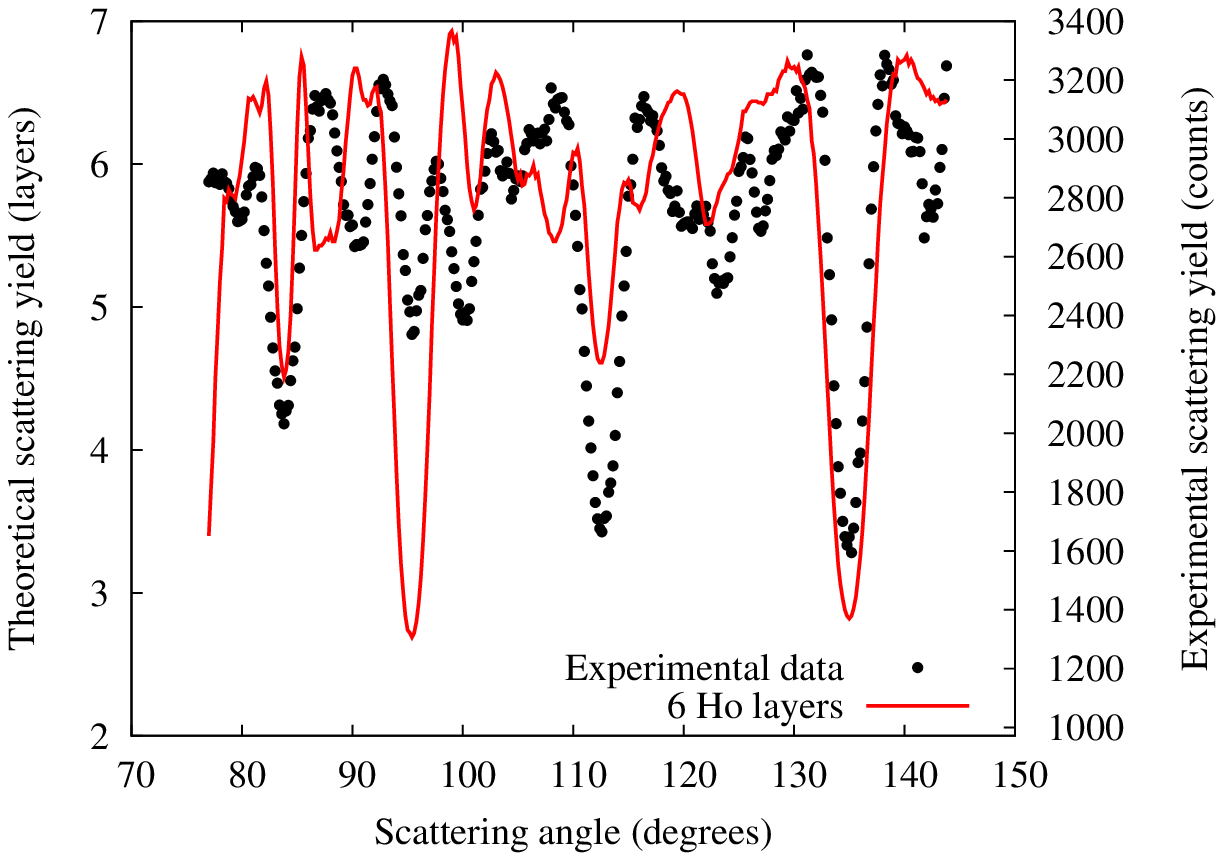}}
    \caption{Original in colour. (a) A comparison of the simulations of the hexagonal structure with 2--6
    layers of Ho, to the experimental data for the \sioneoneo geometry. (b) The fit obtained for 6 Ho layers in the
    hexagonal structure, with a Ho--Ho layer separation of 3.28~\AA.
    }
    \label{fig: hexagonallayerfits}
\end{figure}

Fig.~\ref{fig: hexlayers} shows the simulated blocking curves
obtained when differing numbers of layers of RE are present in the
hexagonal structure for the \sioneoneo geometry. The interlayer
separation of all the Ho layers has been optimised to give the
best visual agreement with experimental data, with the two
intermediate Si layers lying 1/3 and 2/3 of this layer separation
from the Ho layer beneath. A Ho--Ho layer separation of 3.28~\AA\
has been found to give the best-fit. However, the quality of the
fits obtained is very poor, clearly demonstrating that this is not
the correct structure. To make this clearer, the fit for 6 layers
of Ho is shown in Fig.~\ref{fig: hex6layerfit}. At scattering
angles of 88$^{\deg}$, 90$^{\deg}$, 100$^{\deg}$, 106$^{\deg}$,
118$^{\deg}$, 127$^{\deg}$ and 142$^{\deg}$ there are major
discrepancies between the experimental data and the best-fit
simulations. The fully relaxed hexagonal structure that was
proposed by He \etal \cite{He2004A} does not explain the
experimental blocking curves either, as such a structure would
produce almost identical blocking curves, but would yield a
smaller $c$-axis of the silicide.

\begin{figure}[h]
\centering
    \subfigure[]{
    \label{fig: layers110}
    \includegraphics[width=2.9in]{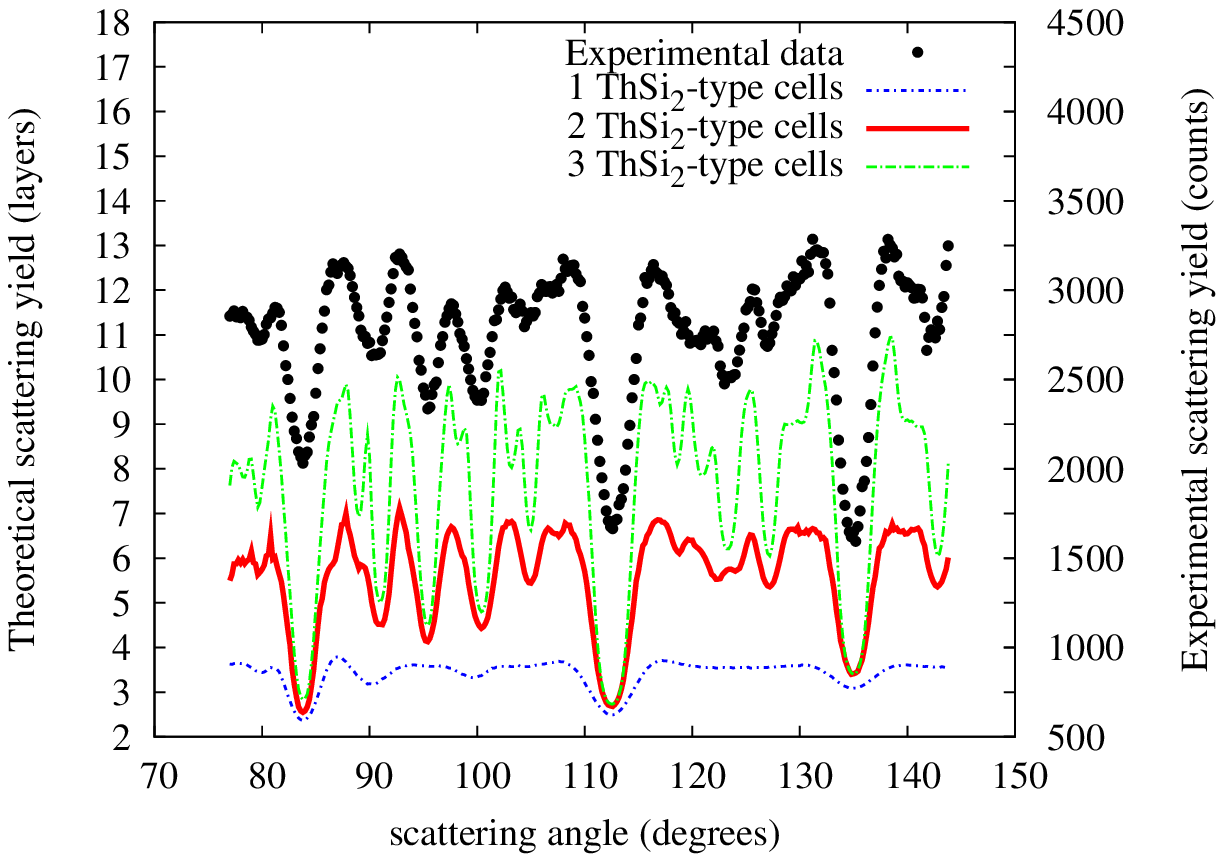}}
    \subfigure[]{
    \label{fig: 8layer110}
    \includegraphics[width=2.9in]{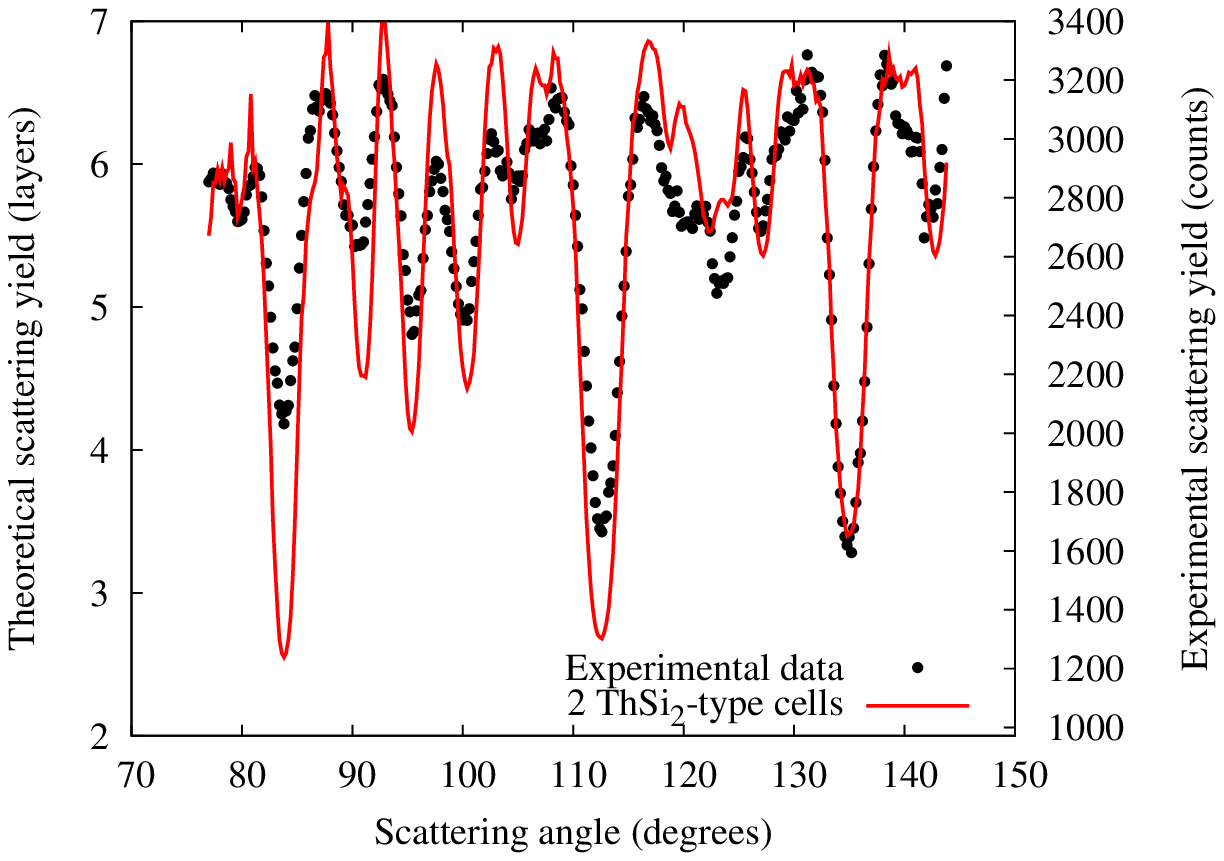}}
    \subfigure[]{
    \label{fig: 8layer111}
    \includegraphics[width=2.9in]{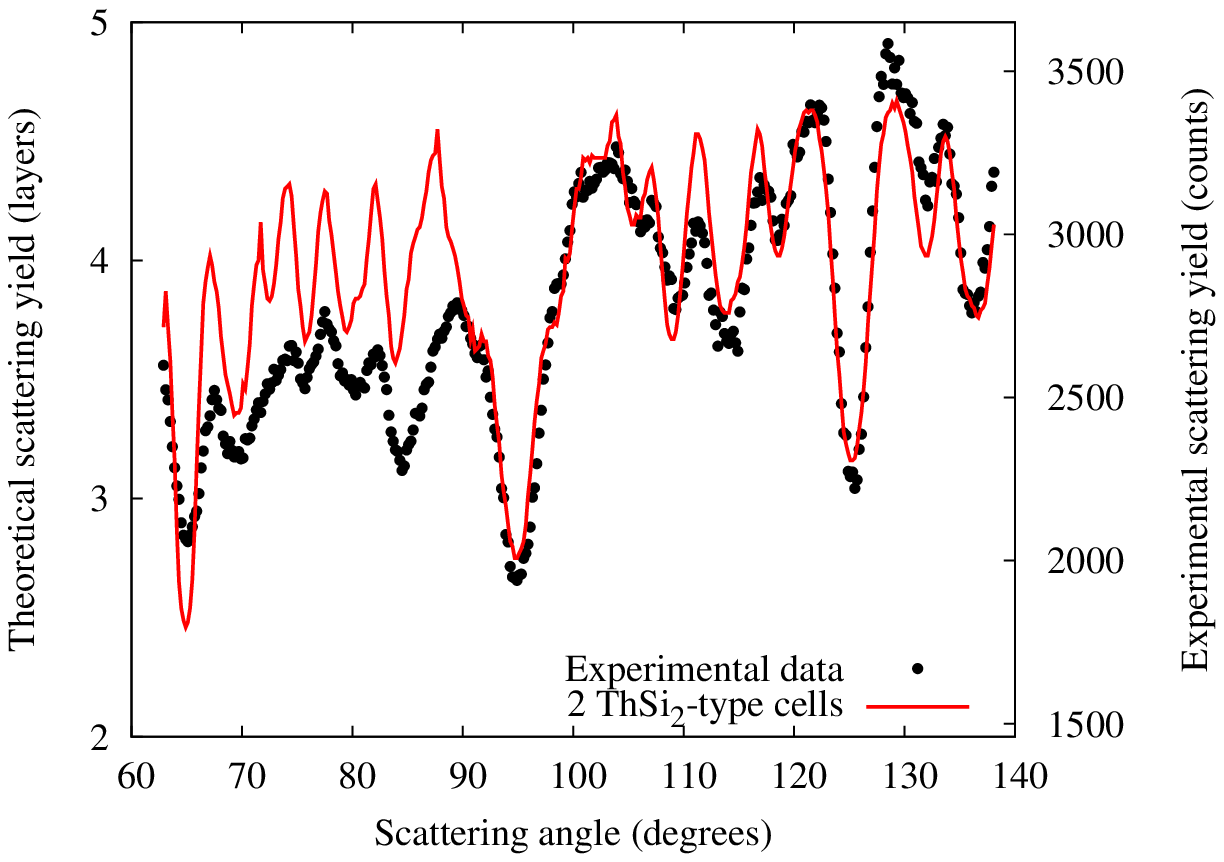}}
    \caption{Original in colour. (a) Comparison of simulations of the tetragonal structure with
    thicknesses of 2--4 ThSi$_2$ cells in height to the experimental data for the \sioneoneo geometry.
    (b) The fit obtained for the \sioneoneo geometry when the silicide is two ThSi$_2$ cells in height.
    (c) The fit obtained for the \sioneoneone geometry when the silicide is two ThSi$_2$ cells in height.}
    \label{fig: tet8layerfit}
\end{figure}

The tetragonal structure has also been fitted to the data, since
it is assumed that the silicide takes the same $a$ and $b$ values
as the Si substrate to form an epitaxial overlayer. This goes
against the available data in the literature regarding the
structure of the \emph{bulk} Ho silicides, as no tetragonal phase
has been observed (though the orthorhombic GdSi$_2$ form has been
reported with lattice parameters of $a$~=~4.03~\AA,
$b$~=~3.944~\AA\ and $c$~=~13.30~\AA) \cite{metalsilicides}.
Conducting simulations of such a tetragonal structure gives the
fits for the \sioneoneo geometry shown in Fig.~\ref{fig:
layers110}, which are for structures that are one, two and three
ThSi$_2$-type cells in height. It is clear that a crystal with a
depth of two ThSi$_2$ cells produces the best match. The detailed
structure fits are shown in figures ~\ref{fig: 8layer110} and
~\ref{fig: 8layer111}. The r.m.s.\ thermal vibrations of the Ho
atoms were set to the bulk metal value of 0.13~\AA, whilst a
fitting of the Si vibrations gives an enhanced value of 0.15~\AA.
These vibrations are consistent with the previous MEIS studies of
2D and 3D RE silicides on Si(111) \cite{Wood2005A,Wood2006A}, with
the factor of two enhancement for the Si atoms possibly indicating
the presence of static disorder in these layers. This could be due
to the presence of Si vacancies \cite{Chi2003A, Tsai2005A}.

Optimising the layer separation of the Ho atoms gives the best
visual agreement for both geometries when $c = 13.14$~\AA. It is
clear from comparing fig.~\ref{fig: 8layer110} with fig.~\ref{fig:
hex6layerfit} that the tetragonal form of the silicide is a better
fit to the data than the hexagonal structure.

\section{AN ANOMALOUS C-AXIS VALUE}

The assumption that the silicide grows epitaxially on the Si(100)
surface yields a constant $c$-axis value throughout this silicide
(13.14 \AA) that is much smaller than the value of 13.30 \AA \,
for the bulk silicide. We would expect that the stress induced
through a contraction of the Ho silicide $a$- and $b$-axes at the
interface, would be released through an expansion in the
perpendicular direction and such behaviour has been directly
observed in other rare earth silicide surfaces
\cite{Bonet2005A,Wood2006A}. In particular, Ye \etal have reported
an expansion in the $c$-axis of the silicide relative to the bulk
value for Dy on Si(100) \cite{Ye2006A}.

One explanation of this observation could be that the silicide is
incommensurate with the Si substrate. Hence it would seem
reasonable to assume that the silicide takes the orthorhombic form
that is observed for bulk Ho silicide. Refitting the experimental
data to simulations of this structure, where $a$ = 4.030~\AA\ and
$b$ = 3.944~\AA, yields almost identical fits, but with a $c$-axis
of 13.70~\AA. This large value of the silicide $c$-axis does not
make sense according to the established trend in strain either, as
it would be expected that the lack of strain at the interface
would not cause any expansion in the $c$-axis of the silicide
relative to the bulk value. Thus it can be concluded that it is
not the bulk form of the silicide that is present on the surface;
it must be some intermediate form.

One of the limitations of MEIS when only using the Ho blocking
curves to solve crystallographic structure is that it is not
possible to exclusively determine both the $a$- and $c$-axis
lattice constants of the silicide. This is because any difference
in the silicide $a$-axis could be compensated by a change in the
$c$-axis to produce blocking curves with blocking dips in the same
positions. Hence, a particular $c/a$ ratio will yield a series of
blocking curves that are exactly the same, even though the actual
lattice parameters may be very different. We have performed a
series of simulations, each with a fixed $a$-axis, in which the
$c$-axis has been optimised by varying the layer separation of a
silicide that is two ThSi$_2$ cells in height. These results are
shown in Table~\ref{tab: aaxisexpand}.

\begin{table}[h]
\begin{center}
\begin{tabular}{ccccccc}  \hline
    \vspace{0.0mm}\\
    $a$ (\AA) & & $c_{110}$ ($\pm$ 0.04 \AA) & & $c_{111}$ ($\pm$ 0.04 \AA) & & $c_{\mathrm{average}}$ ($\pm$ 0.04 \AA) \\
     \hline
    3.84 & & 13.12 & & 13.16 & & 13.14\\
    3.88 & & 13.24 & & 13.28 & & 13.26\\
    3.92 & & 13.40 & & 13.44 & & 13.42\\
    3.96 & & 13.52 & & 13.56 & & 13.54\\
    4.00 & & 13.68 & & 13.72 & & 13.70\\
    \hline
\end{tabular}
\end{center}
\caption{Optimised $c$-axis values determined for various fixed
values of the $a$-axis. Fitting the \sioneoneo and \sioneoneone
geometries yield slightly different values, which can be averaged
to give an overall best-fit value.} \label{tab: aaxisexpand}
\end{table}

\begin{figure}[h]
  \begin{center}
  \includegraphics[width=2.9in]{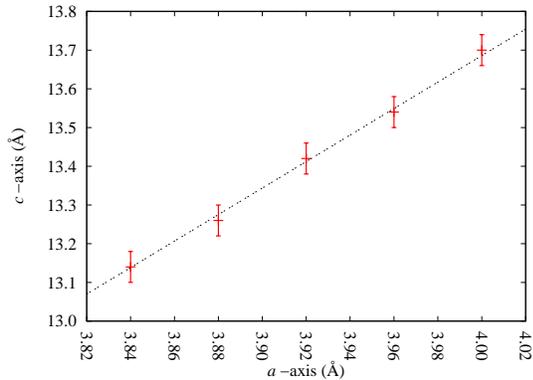}
  \caption{A plot showing how the $c$-axis determined is dependent
  on the size of the $a$-axis of the silicide. Fitting a straight
  line which passes through the origin yields a $c/a$ ratio of 3.42 $\pm$ 0.01.  }
  \label{fig: avsc}
  \end{center}
\end{figure}

The data from Table~\ref{tab: aaxisexpand} are plotted in
Fig.~\ref{fig: avsc} and the straight line fit defines a $c/a$
ratio of 3.42 $\pm$ 0.01. The positive gradient suggests that as
the lateral lattice constant of the tetragonal unit cell is
reduced to match that of the substrate, the lattice constant
perpendicular to the surface also contracts and this is true for
all pairs of $c$, $a$ values. As noted earlier, we might expect
the opposite to occur and the $c$-axis to expand in order to
maintain the volume of the unit cell.

\section{SIMULATION OF STRAINED ISLANDS}

There is another reason that might explain the unusually low
$c$-axis suggested by the fitting procedure. If the lateral
lattice constant of the unit cell is constrained to match that of
the substrate at the interface and is allowed to expand to relieve
strain as we move away from the interface then the blocking angles
will be shifted. The average $a$-axis throughout the structure
will increase and so will the fitted $c$-axis in order to maintain
the $c/a$ ratio of 3.42. This effect would allow for interfacial
matching and would at the same time provide a more physically
meaningful thickness for the silicide.

Simulation of this effect is computationally intensive. Unit cells
with lateral strain as a function of height do not laterally
tesselate \emph{ad-infinitum} and a structure with finite lateral
extent must be considered. This must be large enough to approach
the size of a typical surface island but small enough to be
computationally tractable. We have taken a unit cell with a
10$\times$10 lateral extent, making an island 38.4$\times$38.4
\AA$^{2}$. This is two tetragonal unit cells deep and in total
includes 2500 atoms. Computational resources prohibit us from a
full optimisation of the strained structure in terms of interlayer
spacings or vibrational amplitudes. Instead we have taken values
suggested from the structure fitting carried out on the
1$\times$1$\times$2 cells detailed earlier in this work. The
lateral lattice constants are allowed to relax from a value of
3.84 \AA \,at the bottom of the supercell (interfacial matching to
Si(100)) to 3.99\,\AA \,at the top of the supercell (the averaged
unstrained lateral lattice constant of bulk orthorhombic holmium
silicide). The strain relief is just a linear function of height
and, for example, we have not considered a case in which only the
few layers near the interface are significantly strained. The
strained island is shown from above in Fig.~\ref{fig: strained
island}.

\begin{figure}[h]
  \begin{center}
  \includegraphics[width=8cm]{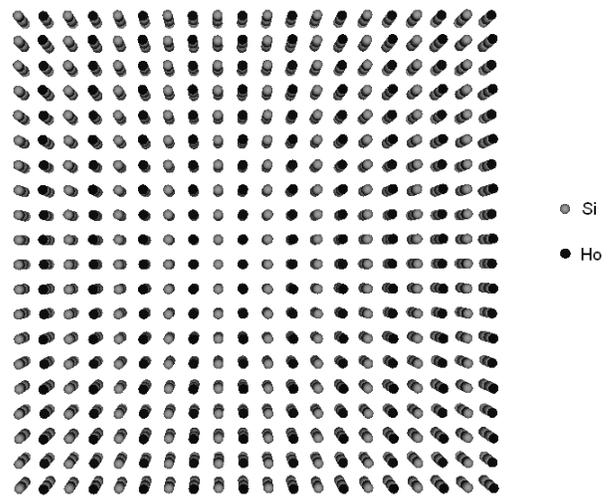}
  \caption{View from above of the 10$\times$10$\times$2
  unit cell showing the lateral strain.}
  \label{fig: strained island}
  \end{center}
\end{figure}

\begin{figure}[h]
  \begin{center}
  \includegraphics[width=9cm]{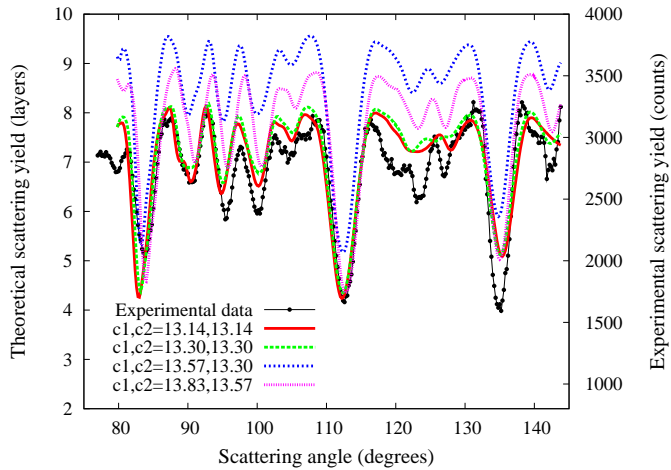}
  \caption{Original in colour. Comparison of simulated strain in large 10$\times$10$\times$2 unit
  cells with experiment. The lateral lattice constant of each unit cell that makes up the nanoisland was allowed to
  relax from 3.84 \AA\, (interface lattice matching with Si(100)) to 3.99 \AA\, on the surface
  (bulk lattice constant). The four simulations plotted here
  correspond to fixed values of the two $c$-axes in the 10$\times$10$\times$2 unit cell.
  The first unit cell (c1) is nearest the surface and the second (c2) is nearest the interface.  }
  \label{fig: strained plots}
  \end{center}
\end{figure}

Since we cannot currently fully optimise the $c$-axis in such a
strained island we have simulated the blocking curves for four
such islands, each with a different depth. This depth is varied by
choosing two different $c$-axes for the two unit cells that form
the island. These four $c$-axis values are; 1) both as bulk
$c$-axis values (13.30 \AA), 2) both at the values obtained
earlier in structure fitting with 1$\times$1$\times$2 cells with
no strain relief (13.14 \AA), 3) expansion of the $c$-axis of the
cell nearest the interface by 2\% and a bulk $c$-axis value in the
top unit cell (13.57 \AA, 13.30 \AA) and 4) expansion of the
$c$-axis of the cell nearest the interface by 4\% and the top cell
by 2\% (13.83 \AA, 13.57 \AA). The calculated blocking curves are
shown in Fig.~\ref{fig: strained plots} for the \sioneoneo
geometry, the \sioneoneone geometry not being calculated because
it would require averaging over two possible domains and thus
twice as many simulations.

On first viewing the incorporation of strain appears to make the
fit worse (compare with the fit for the 1$\times$1$\times$2 cell
in Fig.~\ref{fig: 8layer110}). The region with a scattering angle
in the approximate range 117$^{\circ}$-129$^{\circ}$ is especially
bad. The scattering plane is shown schematically in Fig.~\ref{fig:
scattering_plane}. We can see from this that in the regions where
the fit is bad it is the positions of the silicon atoms that are
responsible for the poor fit of the blocking dips. Another minor
silicon atom blocking dip at around 107$^{\circ}$ is also a poor
fit. Given that the positions and the thermal vibrations of the
silicon atoms have not been optimised it is no surprise that the
fit is made worse by the introduction of strain.

\begin{figure}[h]
  \begin{center}
  \includegraphics[width=8.5cm]{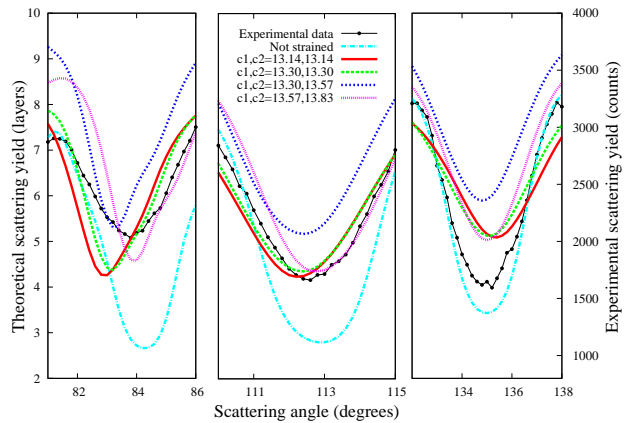}
  \caption{Original in colour. Showing the detail around the major blocking dips in Fig.~\ref{fig:
strained plots}. With the gradual introduction of larger, more
physically reasonable $c$-axis values into the strained cells the
fit is improved until it is better than that for the fully
optimised but non-strained cell.  }
  \label{fig: strain detail}
  \end{center}
\end{figure}

However, the blocking dips produced by this structure are
dominated by those caused by holmium atoms and close inspection of
the major blocking dips at roughly 82$^{\circ}$, 113$^{\circ}$ and
135$^{\circ}$ supports the possibility of strain in these
structures. The detail around these blocking dips is shown in
Fig.~\ref{fig: strain detail}. In the strained cells as the
$c$-axes are changed from the unrealistically low value obtained
in the fit on an unstrained cell (13.14 \AA) through the bulk
silicide value (13.30 \AA) and into the two cases where the
$c$-axis is expanded the blocking dip is shifted to become closer
and closer to the experimentally measured dip. Indeed, at
84$^{\circ}$ and 113$^{\circ}$ the blocking dip for the strained
cells is a better fit to experiment than that for the unstrained,
fully optimised and multilayer averaged fit.

\begin{figure}[h]
  \begin{center}
  \includegraphics[width=8cm]{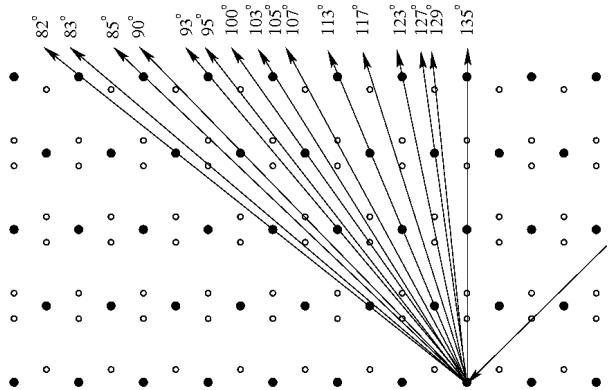}
  \caption{Schematic slice showing the scattering plane for the \sioneoneo
  geometry and the atoms responsible for the principal blocking dips observed.
  Filled circles represent Ho atoms and empty circles silicon atoms.}
  \label{fig: scattering_plane}
  \end{center}
\end{figure}

Thus far, the question of the surface termination has not been
addressed. The unit cells involved in the simulations are holmium
terminated which is not physically reasonable. Also, the LEED
pattern shows a \ctwobytwo periodicity which is not accounted for
in any of the simulated cells. We have attempted to investigate
the surface termination but the simulated blocking curves show
very little sensitivity to an extra layer of silicon atoms that
form a surface termination. Other experiments that are more
sensitive to the very top atomic layer may be better suited to
resolving the issue of the surface termination (such as MEIS
itself at normal incidence with 50 keV He$^{+}$ ions).

\section{Lower Coverages}

\begin{figure}[h]
  \begin{center}
  \includegraphics[width=8cm]{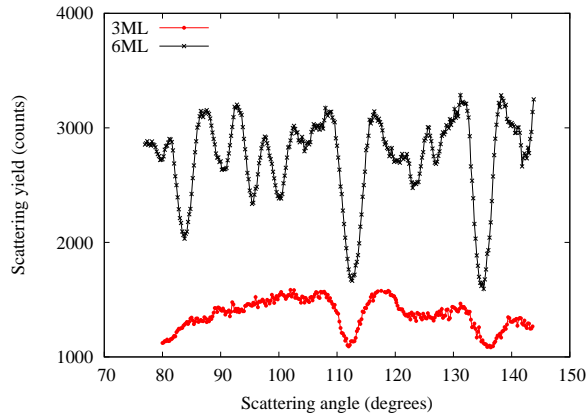}
  \caption{Original in colour. A comparison of the 6 ML coverage experimental data with data recorded after
  a sample was grown under identical growth conditions using a lower
  holmium coverage for the \sioneoneo geometry. The vertical scales have been artificially offset to aid the eye.}
  \label{fig: 3ml}
  \end{center}
\end{figure}

The work presented here suggests that at a coverage of 6 ML and
under our preparation conditions strained tetragonal silicide
islands are formed when holmium is grown on Si(100). At lower
coverages ($<$1 ML) nanowires form that are believed to be related
to the hexagonal structure. The question naturally arises as to
what structure forms in the intermediate coverage regime. Does the
tetragonal structure grow atop the hexagonal phase? Or is there a
phase change at a certain minimum coverage which results in
tetragonal rather than hexagonal growth? To attempt to answer
these question we have taken experimental data at a coverage of
3ML under the same growth conditions. Fig.~\ref{fig: 3ml} shows
these data from the 3 ML sample compared with that from the sample
grown using 6 ML of holmium.

The principal blocking dips at 113$^{\circ}$ and 135$^{\circ}$ are
present in both samples. In the 6 ML sample there are blocking
dips at 90$^{\deg}$, 95$^{\deg}$ and 100$^{\deg}$ that are not
produced from the 3 ML sample. If we refer to Fig.~\ref{fig:
scattering_plane} we can see that these blocking dips are produced
by holmium atoms in the upper region of the structure. We would
not expect to see these features in a thinner silicide layer. In
the hexagonal structure the dominant blocking dip is at a
scattering angle of approximately 95$^{\circ}$ (see fig.
~\ref{fig: hexlayers}). The simulations suggest that this feature
should be present in hexagonal silicides that contain as few as 3
layers of holmium. We can see no evidence for any blocking dips in
this region in the experiment with 3ML coverage and we must
conclude that even at this coverage the structure is tetragonal.

\section{Summary and Conclusion}

Holmium silicide islands have been grown on the Si(100) surface
and characterised using MEIS. Two structures have been fitted to
the experimental data and this process clearly shows that under
our growth conditions it is the tetragonal phase of the silicide
that is formed and not the hexagonal phase and that these two
structures do not coexist on the surface (the orthorhombic phase
is too similar to the tetragonal phase for us to be able to
determine if some of the orthorhombic nature of bulk HoSi is
present in this surface tetragonal phase). A further experiment
using a lower holmium coverage of 3 ML has also been shown to have
the tetragonal structure which confirms that the hexagonal phase
is not the phase adopted at low coverages in this system.

Other authors have reported growth of the hexagonal phase under
similar growth conditions for some of the other RE silicides. It
is interesting to speculate as to why similar experiments have
reached different structural conclusions. It would appear that the
structure that is formed is very sensitive to the growth
conditions. Island morphology is very sensitively dependent upon
the particular RE metal deposited and the annealing temperature
used. In the paper by He et al. \cite{He2004A} they report a
deposition rate of 0.5 ML per minute during the formation of
DySi$_2$ whereas Ye et al. \cite{Ye2006A} report 0.3 ML per
minute. It could be that the mobility of the RE metal and/or the
lattice mismatch anisotropy of the particular silicide in question
are important enough factors during the very early stages of
silicide formation to dictate the final structure.

Using a simple 1$\times$1$\times$2 tetragonal unit cell the
structure fitting suggests a $c$-axis value that is too small to
be physically reasonable when compared to the bulk structure. The
blocking curves from a large 2500 atom 10$\times$10$\times$2
nanoisland with lateral strain relaxation as a function of the
distance from the interface have been simulated. The major
blocking dips produced by this structure are a better fit to the
experimental data than those produced by a non-strained periodic
structure. The lesser blocking dips produced by silicon atoms in
the simulations do not show good agreement with experiment. This
is to be expected because the thermal vibrations and the positions
of these atoms have not been optimised in the structure fit. The
results of the comparison of MEIS data with simulations clearly
show that a physically reasonable $c$-axis value for the holmium
silicide can only be obtained if the islands are strained to fit
the Si(100) at the interface and the strain allowed to relax
towards the top of the islands.

A full structural fit for this system would require an
optimisation of the experiment-theory match with respect to all of
the lateral and vertical spacings in the unit cell and their
variation with distance from the interface, taking into account
lateral strain relief, vertical relaxation and the particular
vibrations of each individual atom. This task is not currently
tractable in terms of computational resources and the available
computer codes.

\begin{acknowledgements}

The authors would like to acknowledge the Engineering and Physical
Science Research Council for funding this research. Paul Quinn and
the FOM Institute are thanked for supplying the XVegas and Vegas
simulation codes. The assistance and technical support of Kevin
Connell and Mark Pendleton are also much appreciated.

\end{acknowledgements}

\end{document}